# On the Anticipation of Lunar Travel in the Early 20th Century: A Pedagogical Exercise[1]

Tina A. Harriott,*[a] Chérif F. Matta*[b]

[a] Department of Mathematics and Statistics; [b] Department of Chemistry and Physics, Mount Saint Vincent University, Halifax, Nova Scotia, Canada B3M 2J6.
Tina.Harriott@msvu.ca; cherif.matta@msvu.ca

## GRAPHICAL ABSTRACT

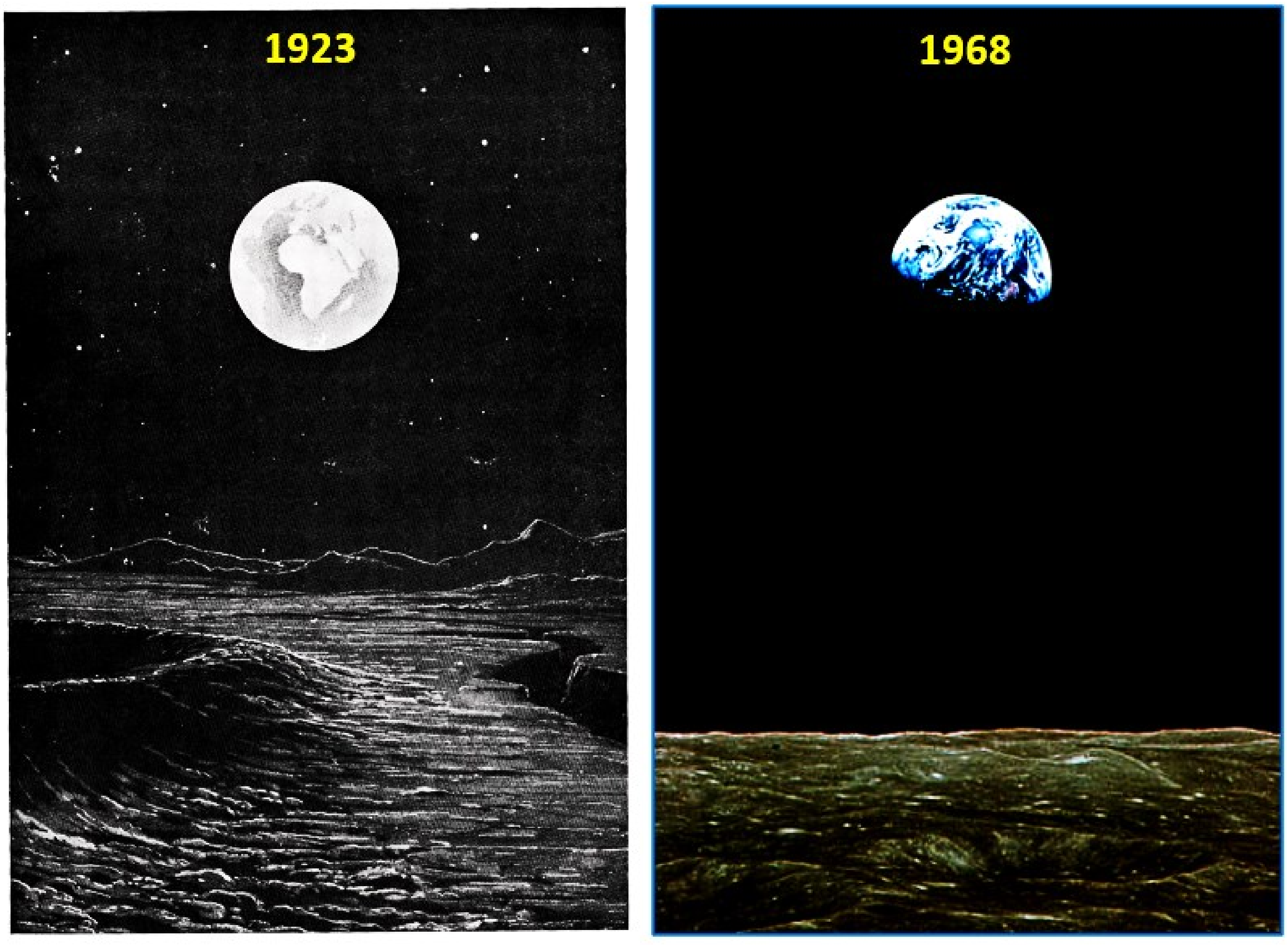

[1] C.F.M. dedicates this article to the memory of his parents, Farid A. Matta and Nabila N. Abdel-Nour, whose well-stocked family home library in Alexandria introduced him, in his childhood, to the beautiful book of Alphonse Berget, *Le Ciel*. This book has profoundly influenced C.F.M.'s lifelong trajectory in science and interest in astronomy.

**ABSTRACT**

This article examines, from historical and pedagogical perspectives, Alphonse Berget's anticipation of Earth–Moon travel in Le Ciel (Larousse, 1923), decades before the beginning of the space age. The discussion is triggered by *Le Ciel*, a richly illustrated French popular science work, which has a devoted chapter examining lunar and interplanetary travel within a Newtonian framework. Although Berget's treatment was not developed in isolation and reflects a broader early 20$^{th}$ century context that included pioneers such as French aero-engineer Robert Esnault-Pelterie, the book provides a striking pedagogical synthesis of elementary celestial mechanics and scientific popularization.

Unlike earlier fictional treatments such as Jules Verne's *De la Terre à la Lune*, Berget approached space travel using physical reasoning grounded in Newtonian gravitation. Using qualitative and semi-quantitative arguments based on the inverse-square law, he identified the principal phases of an Earth–Moon trajectory: escape from Earth, inertial translunar motion, transition through competing Earth–Moon gravitational fields, and final lunar capture and deceleration. His estimated Earth–Moon travel time of approximately 49 hours is of the same order of magnitude as Apollo mission transit times (~72 h).

We compare these early ideas with modern elementary concepts of astrodynamics, including restricted three-body trajectories, Lagrange-point dynamics, and distant retrograde orbits associated with the Artemis program. We also examine Berget's discussion of interplanetary travel, lunar landscapes, and human factors associated with prolonged voyages, including confinement, food supply, and travel duration. The analysis highlights the pedagogical value of historically grounded scientific reasoning underpinning spaceflight mechanics.

**KEYWORDS** Robert Esnault-Pelterie. Alphonse Berget. *Le Ciel*. History of astronautics. Lunar travel. Astrodynamics. Restricted three-body problem. Interplanetary travel. Pedagogy of astrophysics. Popularization of Science.

## 1. When Physics Precedes Technology

There are rare moments in the history of science when the conceptual structure of a technological achievement is laid out long before its realization.

Konstantin Tsiolkovsky (1857–1935) is often called the grandfather of rocketry. He was a Russian schoolteacher who in a 1903 work titled *Exploration of Outer Space by Means of Rocket Devices* included the now so called "rocket equation", proposed liquid propellants, and suggested multi-stage rockets [1]. Indeed, more than 50 years later and 20 years after Tsiolkovsky's death, Russia was the first country to successfully launch a spacecraft (Luna 1) that escaped Earth's gravity, on January 2, 1959 [2].

The case of lunar travel is often associated with 20th century pioneers such as Robert Goddard, who is credited with creating and building the world's first liquid fueled rocket, successfully launched in 1926 [3,4] and Wernher von Braun, who headed the team that developed the Saturn V rocket in the 1960's that ultimately carried the Apollo astronauts on their lunar missions [5]. These advances in rocketry along with the development of digital computers, and use of orbital mechanics led to the realization of human travel to the Moon, yet its underpinnings are entirely contained within classical Newtonian physics.

A notable popularizing review of the state of knowledge of the time by Alphonse Berget in his 1923 beautifully illustrated monograph "*Le Ciel* (The Sky)" [6], published by Larousse in Paris (Fig. 1), shows that the idea of travel to the Moon is already discussed in a qualitative but physically sound manner almost half a century before humans set foot on the Moon in 1969.

Berget credits his compatriot Robert Esnault-Pelterie (1881-1957), Fig. 2, and uses his predictions, not in speculative manner of fiction but by using solid mathematical and physical arguments. His discussion begins with the most elementary principle, and that is, the inverse square law of gravity. From this, he constructs, not formally, but unmistakably, the phases and milestones of a lunar voyage from Earth with humans inside what he calls a "projectile". To estimate the total time of the Earth–Moon trip, Berget suggests [6]:

> " La durée totale du voyage peut se faire par addition des durées des trois phases…"
>
> *"The total duration of the journey can be obtained by adding the durations of the three phases…"*

This is not merely descriptive. It is the *first step of a dynamical decomposition* implying that motion must pass through qualitatively distinct regimes. That insight alone places Berget beyond the speculative imagination and within the domain of sound physical reasoning.

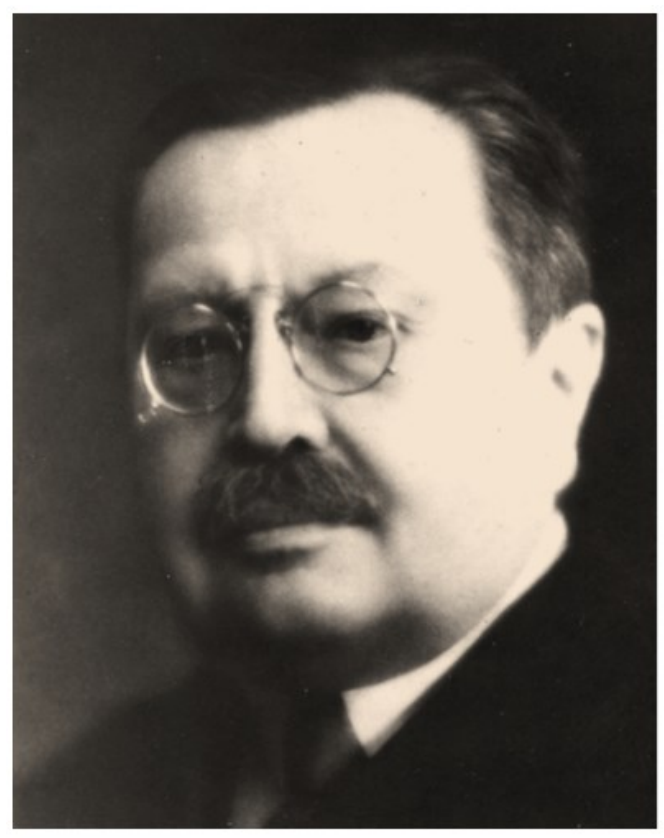

(Baron) Alphonse Berget
(1860 – 1933)

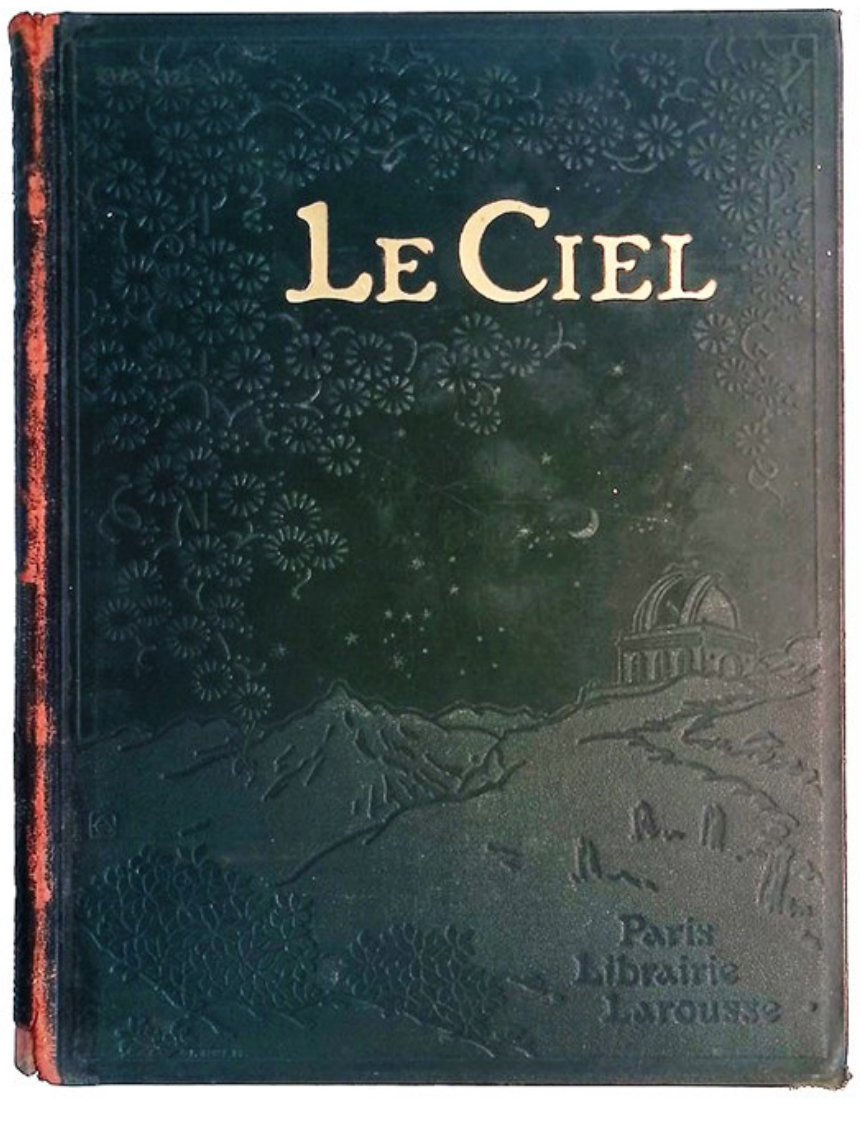

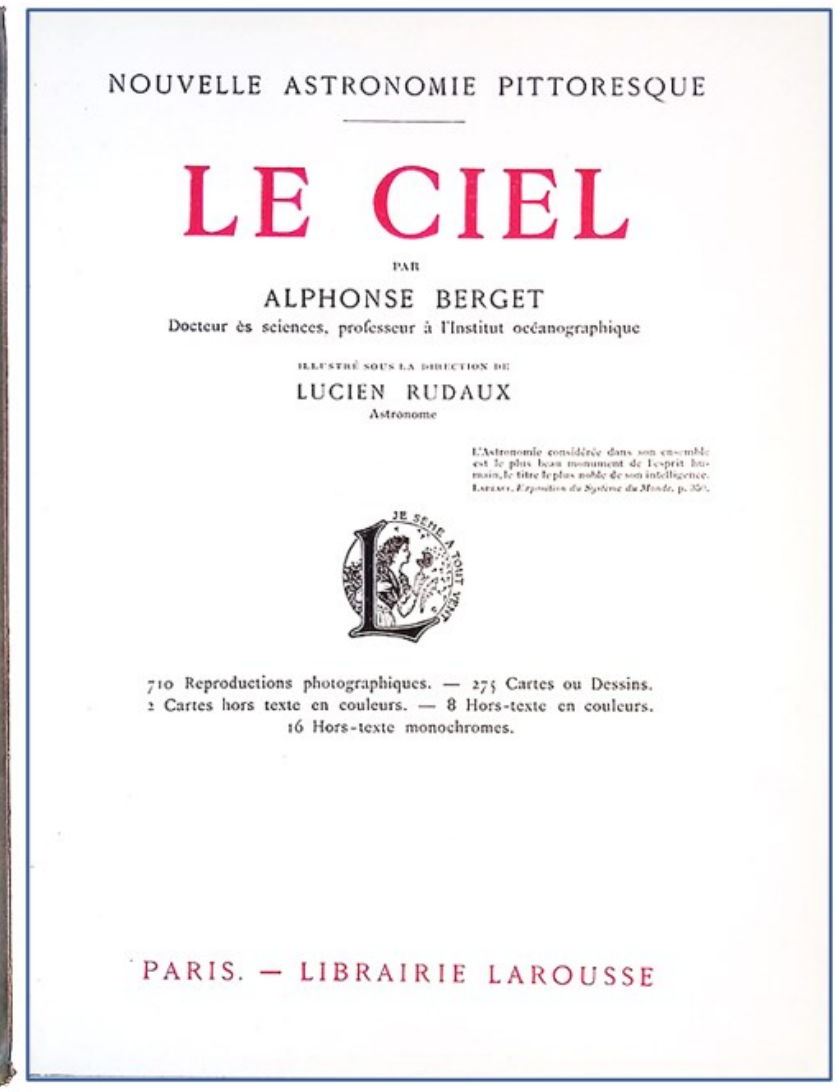

**Figure 1.** Portrait of Alphonse Berget (left), together with the cover and title page of his book *Le Ciel* (centre and right) [6]. Berget (1860–1933) was a French science writer and educator associated with the Larousse publishing house, known for synthesizing contemporary astronomical (and broader scientific) knowledge for a broad audience. Trained in the sciences and active in early 20th-century scientific communication, he combined rigorous physical reasoning with pedagogical beautifully illustrated exposition. In *Le Ciel* (1923), he uses Newtonian dynamics to describe the conditions and phases of a lunar travel decades before their technological realization. (Public domain).

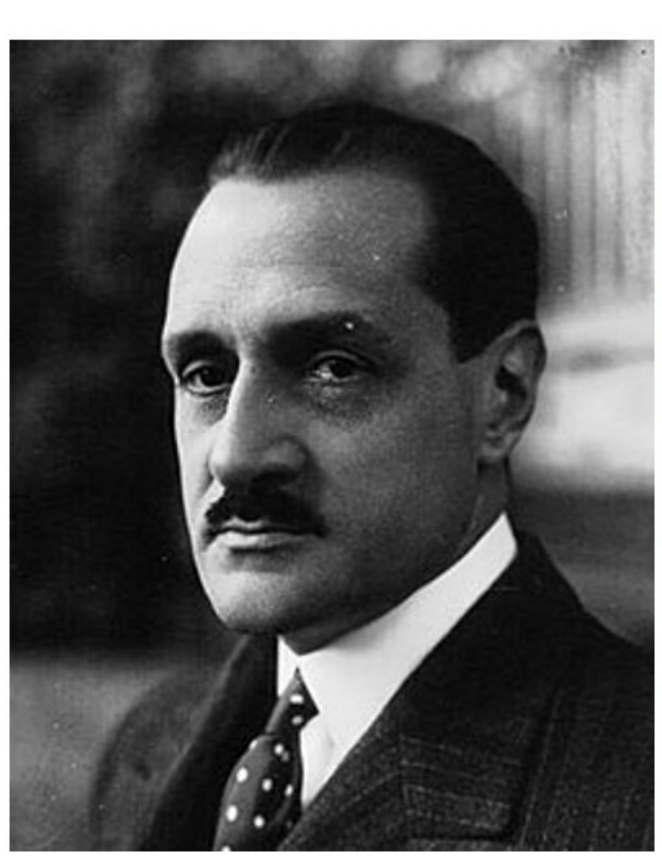

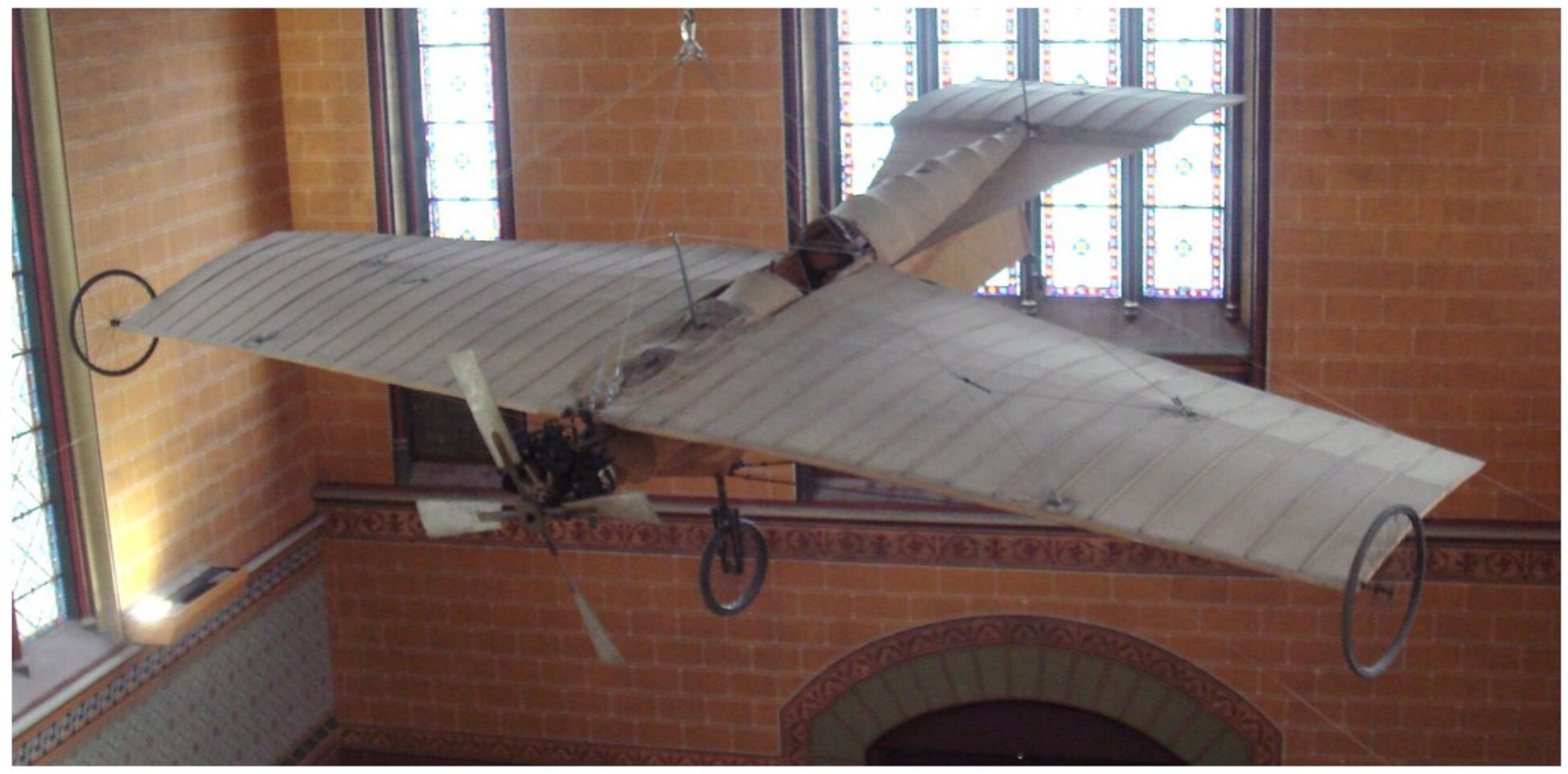

**Figure 2.** Robert Esnault-Pelterie (1881-1957) (left) and an early aircraft design associated with his work (right). Esnault-Pelterie was a pioneering French aviator and engineer who made significant contributions to early aviation and was among the first to formulate a theoretical framework for space travel, including independent derivations of the rocket equation in the 1910s. His work predates and informs later discussions of astronautics in the early 20th Century. Image source: Wikipedia (public domain / Wikimedia Commons).

## 2. From the Fiction of Jules Verne to the Dynamics of Alphonse Berget

A little over a century before the Apollo lunar missions, Jules Verne, in 1865, published the fiction *De la Terre à la Lune* (From the Earth to the Moon) [7] – Fig. 3. Berget summarizes the thesis of Jules Verne in a section that he starts (on page 180) by stating [6]:

> "Une première méthode a été rendue populaire par le prestigieux roman de Jules Verne, *De la Terre à la Lune*."
>
> "*A first method was popularized by Jules Verne's celebrated novel* From the Earth to the Moon."

Verne's novel is celebrated for its pre-science as it is anchored in a pre-dynamical concept whereby the journey is reduced to a single act, that of the launch. In this context, in Chapter VII of the novel, Verne writes:

> "...il résulte de mes calculs indiscutables que tout projectile doué d'une vitesse initiale de douze mille yards par seconde, et dirigé vers la Lune, arrivera nécessairement jusqu'à elle."
>
> *"...it results from my indisputable calculations that any projectile endowed with an initial velocity of twelve thousand yards per second, and directed towards the Moon, will necessarily reach it."*

Berget summarizes Jules Verne's method of lunar travel as the launch of a projectile from Earth with sufficiently high initial velocity to carry passengers toward the Moon. The journey is conceived essentially as a ballistic motion initiated by a single powerful impulse, after which the projectile continues through space largely under its acquired momentum. Berget recognizes the imaginative power of this idea but implicitly contrasts it with a more physically grounded treatment based on the continuous action of gravitational forces throughout the voyage. While Verne emphasizes the launch itself, Berget shifts the focus toward the successive dynamical regimes governing the entire Earth–Moon trajectory.

Jules Verne's approach is one totally determined by the initial velocity ($v_0$) condition:

$$v(t = 0) = v_0, \tag{1}$$

after which the motion is largely "taken for granted". There is no analysis of the governing equation:

$$\frac{d^2r}{dt^2} = -\frac{GM}{r^2}. \tag{2}$$

where $M$ is the mass of the Earth, $r$ is the (changing) separation of the rocket and the center of the Earth and $G$ the universal constant of gravitation ($G = 6.67 \times 10^{-11}\mathrm{m}^3\mathrm{kg}^{-1}\mathrm{s}^{-2}$). Neither is there any discussion of the growing influence of the Moon's gravity as it is approached. In Verne, gravity is present but not structurally organizing. In Berget-Esnault-Pelterie, it becomes the central driving force.

In *Le Ciel* the idea of a single decisive impulse is replaced with a *continuous evolution under force*, reflecting the prevalent view in the 1920's. This is a conceptual shift of first importance. It transforms lunar travel from a single event into an evolving phased process.

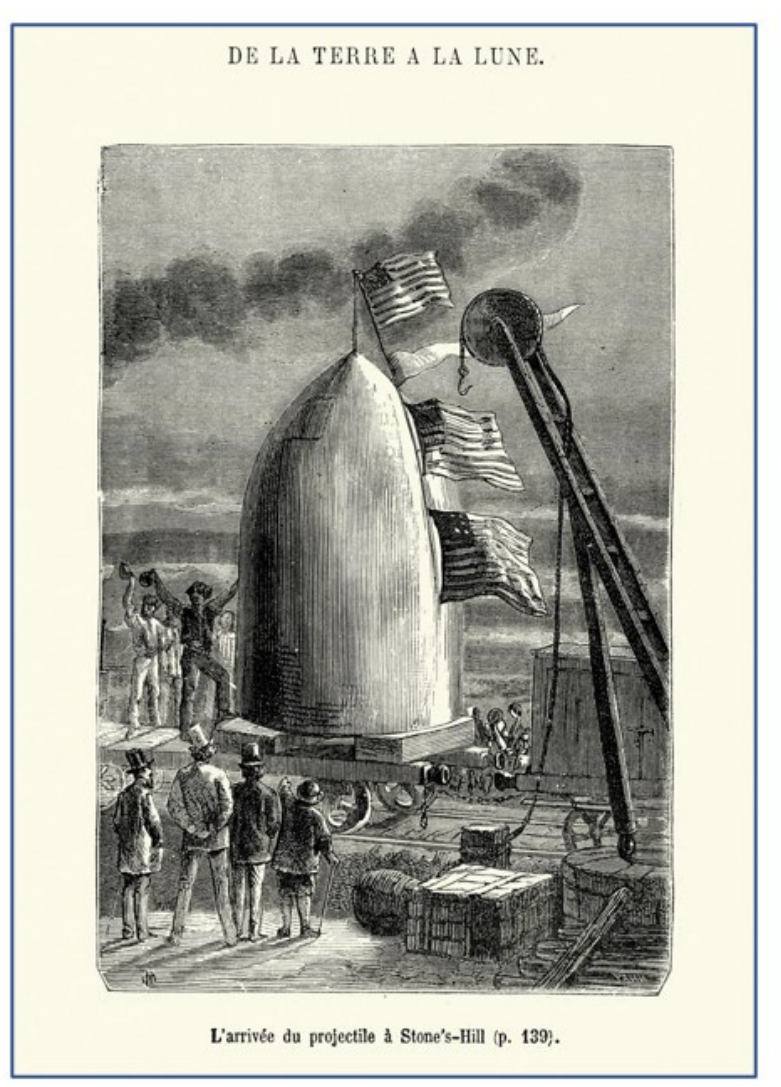

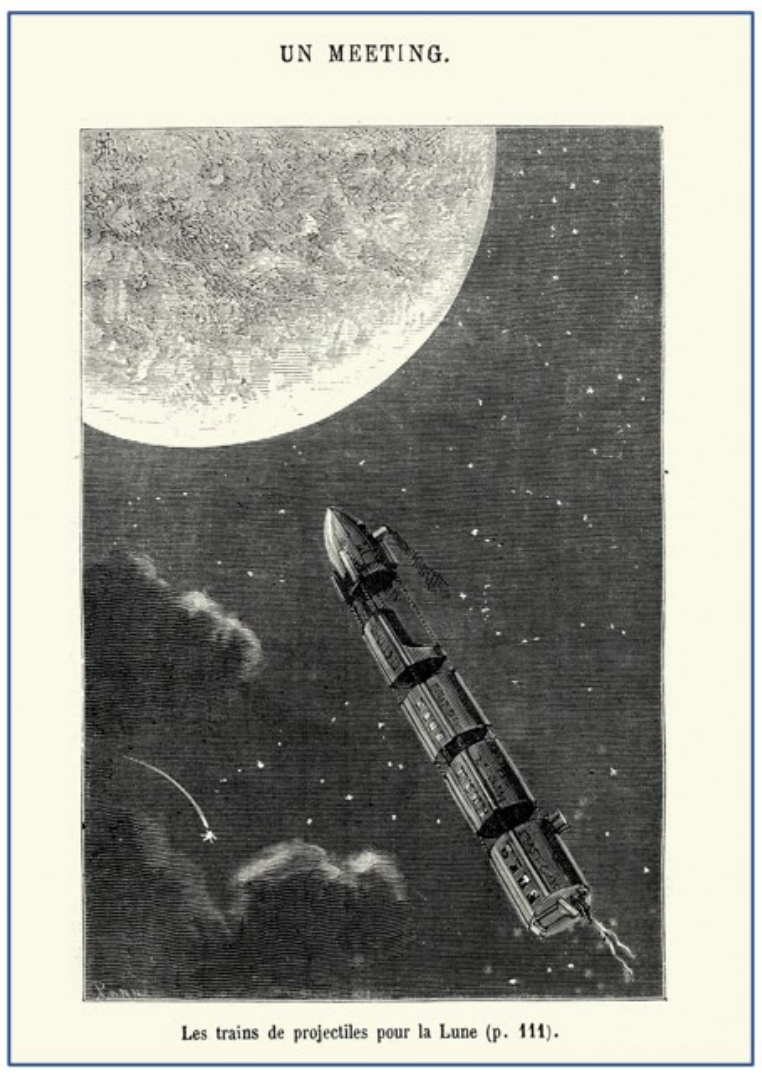

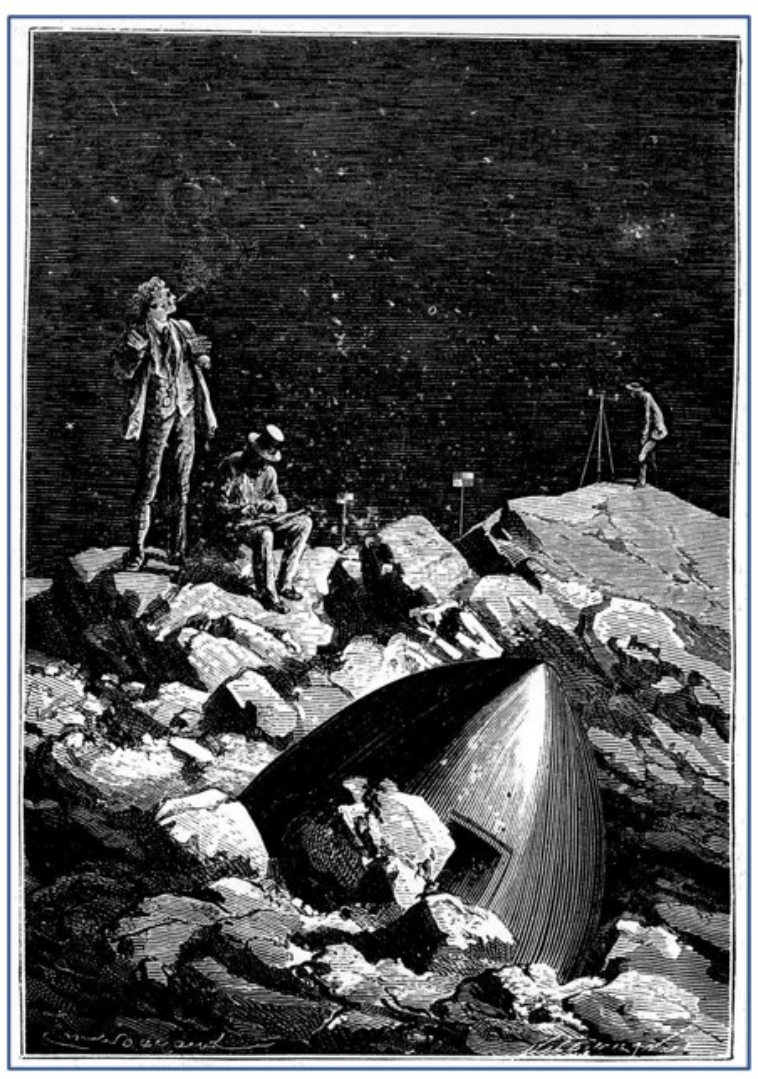

**Figure 3.** Illustrations from 1865 novel *De la Terre à la Lune* (from the Earth to the Moon) and its sequel by Jules Verne [7]. Left: preparation and emplacement of the projectile at Stone's Hill, launched by cannon toward the Moon. Centre: the projectile in flight toward the lunar surface. Right: imagined arrival and exploration on the Moon. Verne's depiction assumes an impulsive launch with large initial velocity and treats the trajectory as essentially ballistic, without a continuous dynamical analysis of motion under gravitational forces. These images illustrate an early conceptual stage of lunar travel, preceding the physically grounded treatment developed and summarized nicely by Berget. (Public domain).

## 3. Escape from Earth: The Energy Barrier

The first physical obstacle for travel from the Earth to the Moon is the gravitational attraction of Earth on the object undergoing the travel. The total energy of a body of mass *m* in Earth's gravitational field, in standard notation, is:

$$E = \frac{1}{2}mv^2 - \frac{GM_{\oplus}m}{r}. \tag{3}$$

where $m$ is the mass of the object undergoing the travel and $M_{\oplus}$ is the mass of the Earth. That is, the total energy of the object is given by the sum of its kinetic energy and its gravitational potential energy (due to the Earth's gravitational field).

Escape requires that this energy be non-negative. Setting $E = 0$ at Earth's surface yields

$$v_{\text{esc}} = \sqrt{\frac{2GM_{\oplus}}{R_{\oplus}}}. \tag{4}$$

where the separation at the Earth's surface is $r = R_{\oplus}$, where $R_{\oplus}$ is the radius of the Earth. Numerically, this translates into $v_{\text{esc}} \approx 11.2$ km/s.

While Berget does not present this equation explicitly, he explains that a sufficiently large initial velocity is required to overcome Earth's attraction. The presence of the concept is what is significant, and not the absence of the formula. Berget recognizes that the problem is fundamentally energetic, *i.e.*, that the spacecraft must climb out of a potential well. This language, though not formalized, is entirely consistent with modern treatments.

## 4. The Decay of Gravitational Influence and the Take-Over by Inertia

Once the spacecraft has departed Earth, the magnitude of the gravitational force, $F(r)$, between the Earth and the object, governing its motion decreases rapidly:

$$F(r) = \frac{GM_{\oplus}m}{r^2}. \tag{5}$$

The ratio

$$\frac{F(r)}{F(R_{\oplus})} = \left(\frac{R_{\oplus}}{r}\right)^2 \tag{6}$$

reveals the scale of this decay. At ten Earth radii, the influence of Earth's gravity is already reduced by two orders of magnitude.

What follows from this is crucial. As the force weakens, acceleration tends to zero:

$$\frac{d^2r}{dt^2} \to 0, \tag{7}$$

and thus its velocity $v \to$ constant.

Berget does not state this explicitly, but his narrative implies it by emphasizing that the spacecraft's motion becomes essentially inertial. This corresponds directly to the translunar coast phase in Apollo missions. At this phase, the spacecraft is no longer "driven" outward; it drifts. This is a profound shift in physical regime, and Berget captures it without formalism.

## 5. The Gravitational Competition: Earth *vs.* Moon

The analysis reviewed qualitatively in *Le Ciel* recognizes that there exists a region where Earth and Moon exert comparable influences. This is not obvious. It requires understanding that gravitational forces from different bodies can compete in space.

The condition for equality in the magnitude of these forces (which are opposite in direction) is that the size of the Earth's gravitational attraction equals the Moon gravitational attraction, *i.e.*,

$$\frac{GM_{\oplus}}{r^2} = \frac{GM_{\text{Moon}}}{(d-r)^2}, \tag{8}$$

where $r$ is the distance from the center of the Earth to the point where the magnitudes of the gravitational attractions of Earth and Moon become equal (the "balance point"), $d$ = Earth–Moon distance ($d \approx 3.84 \times 10^8\ m$), and, clearly, ($d$–$r$) is simply the distance from this balance point to the center of the Moon. Solving for $r$ we get:

$$r = \frac{d}{1 + \sqrt{\frac{M_{\text{Moon}}}{M_{\oplus}}}}. \tag{9}$$

Using $M_{Moon} = 7.3477 \times 10^{22}$kg, and $M_{\oplus} = 5.9722 \times 10^{24}$ kg, one obtains

$$\sqrt{\frac{M_{\text{Moon}}}{M_{\oplus}}} \approx 0.11, \tag{10}$$

and hence $r \approx 0.90\ d$.

Berget argues that there exists a region where the two attractions nearly compensate one another, anticipating the modern notion of gravitational boundaries and, more broadly, the three-body problem, yet without invoking concepts such as Hill spheres which are the regions around a celestial body (such as the Earth) within which its gravity is the dominant force governing the motion of smaller objects (such as a rocket launched from its surface).

## 6. Capture by the Moon: A Second Dynamical Regime

Beyond the transition region, the Moon becomes dominant. The equation of motion changes character, and the spacecraft acceleration is now directed primarily toward the Moon:

$$\frac{d^2\mathbf{r}}{dt^2} = -\frac{GM_{\text{Moon}}}{|\mathbf{r}|^2}\hat{\mathbf{r}}, \tag{11}$$

where $\mathbf{r}$ is the spacecraft position vector relative to the Moon, and $\hat{\mathbf{r}}$ is the radial unit vector directed away from the Moon.

The spacecraft is now in a new gravitational potential well. To remain bound to the Moon, its mechanical energy relative to the Moon must satisfy

$$E_{\text{Spaceccraft}} = \frac{1}{2}mv^2 - \frac{GM_{\text{Moon}}m}{r} < 0. \tag{12}$$

where $r$ is the spacecraft distance from the Moon and $v$ its total speed relative to the Moon (the magnitude of the spacecraft velocity vector $\mathbf{v}$ in the Moon-centered frame). This condition implies that the spacecraft velocity must be sufficiently reduced upon approaching the Moon in order for the total mechanical energy to become negative. In practice, this requires a deceleration maneuver to prevent the spacecraft from simply passing by the Moon on an unbound trajectory.

Berget notes qualitatively that the velocity would need to be adjusted near the Moon. In modern astronautics, this corresponds to the lunar orbit insertion maneuver.

## 7. A Qualitative Comparison with the Apollo missions: Similarities and Limits

Apollo missions provide a qualitative benchmark for assessing the staged structure of the lunar voyage described in *Le Ciel*, while also revealing the many elements absent from Berget's popularizing review. According to the author, the voyage begins with the *initial powered phase*, during which the projectile is accelerated to a very high velocity of approximately 11,800 m/s by Eq. (4). This value is close to modern escape velocity (≈ 11,200 m/s, see above). This phase is short in duration, lasting, according to the author, only *ca*. 29 minutes, corresponding to overcoming Earth's gravitational attraction.

The motion then enters a *long inertial (coasting) phase*, during which the projectile travels under the combined influence of Earth and Moon gravity. Berget notes that the velocity gradually decreases to a minimum near the region where the attractions of the two bodies balance, after which it increases again as the projectile falls toward the Moon. This phase dominates the total travel time and is estimated to last about 48.5 hours.

Finally, Berget introduces a *third phase of deceleration ("freinage" (braking))*, required to ensure that the projectile reaches the lunar surface with a manageable velocity. Without this braking, the impact velocity would remain high. He estimates a short braking interval of a few minutes, during which the speed is reduced sufficiently for landing.

A particularly instructive passage in Chapter XIII of *Le Ciel* follows the decomposition of the Earth–Moon journey into the above-mentioned phases the prelude of which reads (p. 183) [6]:

> "La durée totale du voyage peut se faire par addition des durées des trois phases…"
>
> *"The total duration of the journey can be obtained by adding the durations of the three phases…"*

followed immediately by the following estimates [6]:

| | |
|---|---|
| Première phase / *First phase*: | 0 h 24 m 9 s |
| Deuxième phase / *Second phase*: | 48 h 30 m |
| Troisième phase / *Third phase*: | 0 h 3 m 46 s |
| Durée totale / *Total duration*: | 48 h 58 m (environ / *approximately*) |

The above tabulation explicitly identifies three regimes (acceleration, inertial motion under gravity, and terminal deceleration) that correspond precisely to the modern phases of translunar flight. Furthermore, the total duration is of the correct order of magnitude, despite the absence of any formal orbital mechanics. Accordingly, the total duration of the journey, obtained by summing these three phases, is of roughly 48 hours 58 minutes ($t_{\mathrm{Berget}} \approx 49$ h), remarkably close in order of magnitude to modern translunar flight times ($t_{\mathrm{Apollo}} \approx 72$ h) with an error of only −32%.

The Apollo missions provided the first empirical realization of lunar travel. Their trajectories followed a canonical sequence consisting of launch into Earth orbit, translunar injection, translunar coast, lunar orbit insertion, descent, and landing [8−10].

This sequence corresponds approximately to the phases described in *Le Ciel*. The primary difference lies in the mathematical description. Apollo trajectories are governed by orbital mechanics, in which the motion is described by conic sections and characterized by quantities such as the semi-major axis $a$, of an ellipse, with the total mechanical energy, $E$, of the spacecraft of mass, $m$, given by [11,12]:

$$E = -\frac{GMm}{2a}. \tag{13}$$

where $M$ is the mass of the central body (Earth or Moon). The equation is that of the well-known bound Keplerian orbit and obtained from the sum of kinetic and gravitational potential energies averaged over the orbit and satisfying the virial theorem (which states that potential energy is equal to twice the negative of the kinetic energy averaged over the elliptical orbit's period).

For typical spacecraft of mass $m$, $m \ll M$, so the motion, to a very good approximation may be treated as that of a test particle orbiting a fixed central mass. Thus, Berget's success arises from the correct use of inverse-square law, scaling arguments, and, importantly, the decomposition of motion into phases. It is limited, however, by the absence of detailed orbital mechanics formalism, *e.g.*, Eq. (13), the absence of propulsion constraints, and the lack of an explicit discussion of angular momentum:

$$\mathbf{L} = \mathbf{r} \times \mathbf{p}, \tag{14}$$

and, therefore, cannot describe orbital motion. These limitations are central and prevent Berget's account from being treated as a modern astro-dynamical analysis; its value is instead pedagogical and historical.

In 1923, aviation was still in its formative stage. Aircraft were fragile biplanes (Figs. 2 and 4) with limited range, low speed, and minimal reliability. Commercial air travel, though initiated in a few isolated and short routes, had not yet developed into a stable or widespread system. Typical flight speeds of order $10^2$ km/h were *two orders of magnitude below the velocities required for escape from Earth's gravitational field*, $v_{esc} \approx 40{,}000$ km/h. Under these conditions, the very idea of systematic travel beyond the atmosphere remained technologically remote.

It is in this context the early 20$^{th}$ Century leaps of imagination captured in Berget's book must be appreciated. At a time when sustained intercontinental flight was not yet operational and intercontinental travel was by sea, he approached lunar travel as a concrete real future possibility governed by Newtonian mechanics and correctly identified in qualitative form several physically meaningful stages, their duration, and structure.

The contrast with the Saturn V launch vehicle of the Apollo program, shown in Fig. 4, underscores the magnitude of this leap. By 1968, propulsion technology had advanced to the point where velocities of several kilometers per second could be achieved, enabling translunar injection and controlled spaceflight.

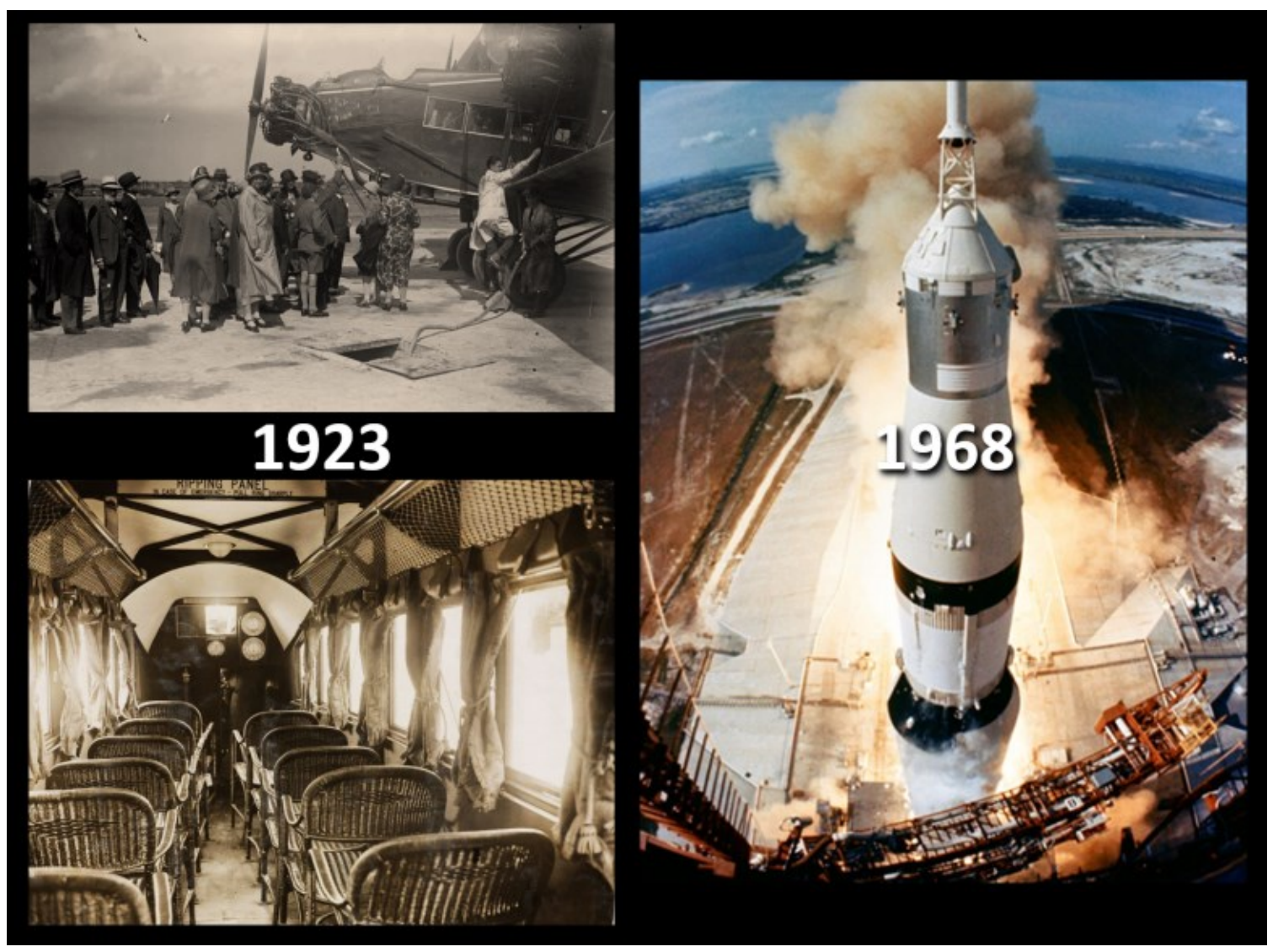

**Figure 4.** Contrast between the state of aviation in 1923 and the launch capability achieved during the Apollo program. Left: aviation context in the early 1920s. Top: passengers and crew boarding a propeller-driven aircraft (*ca.*

1920s). Bottom: interior of an early commercial aircraft cabin, illustrating limited capacity and rudimentary comfort. Right: launch of the Saturn V rocket (Apollo era, late 1960s), capable of delivering payloads to translunar trajectories with velocities approaching escape velocity. The juxtaposition highlights the gap between early aviation and spaceflight. (Left images: public domain (early aviation photography, *ca.* 1920s). Right image: NASA, Saturn V launch (public domain)).

In *Le Ciel* we find an imagined view of the Earth as seen from the lunar surface, rendered with a realism that is remarkable for its time. The terrestrial globe is depicted with a large apparent angular diameter, suspended in a black sky devoid of atmospheric scattering, and set above a stark, high-contrast landscape. A comparison of the diameters of the Earth seen from the Moon and the Moon seen from the Earth, displayed side by side on page 184 of the book, reveals that the angular diameter of the Earth as seen from the Moon is approximately $\theta \approx 2^{o}$, about four times that of the Moon seen from Earth. The absence of atmosphere ensures sharp contrasts between illuminated and shadowed regions as opposed to the Earthly landscape juxtaposed to it (Fig. 5).

The famous Apollo 8 "Earthrise" image and the Apollo 17 views of Earth above the lunar horizon confirm the same geometric and photometric features anticipated in *Le Ciel* half a century earlier.

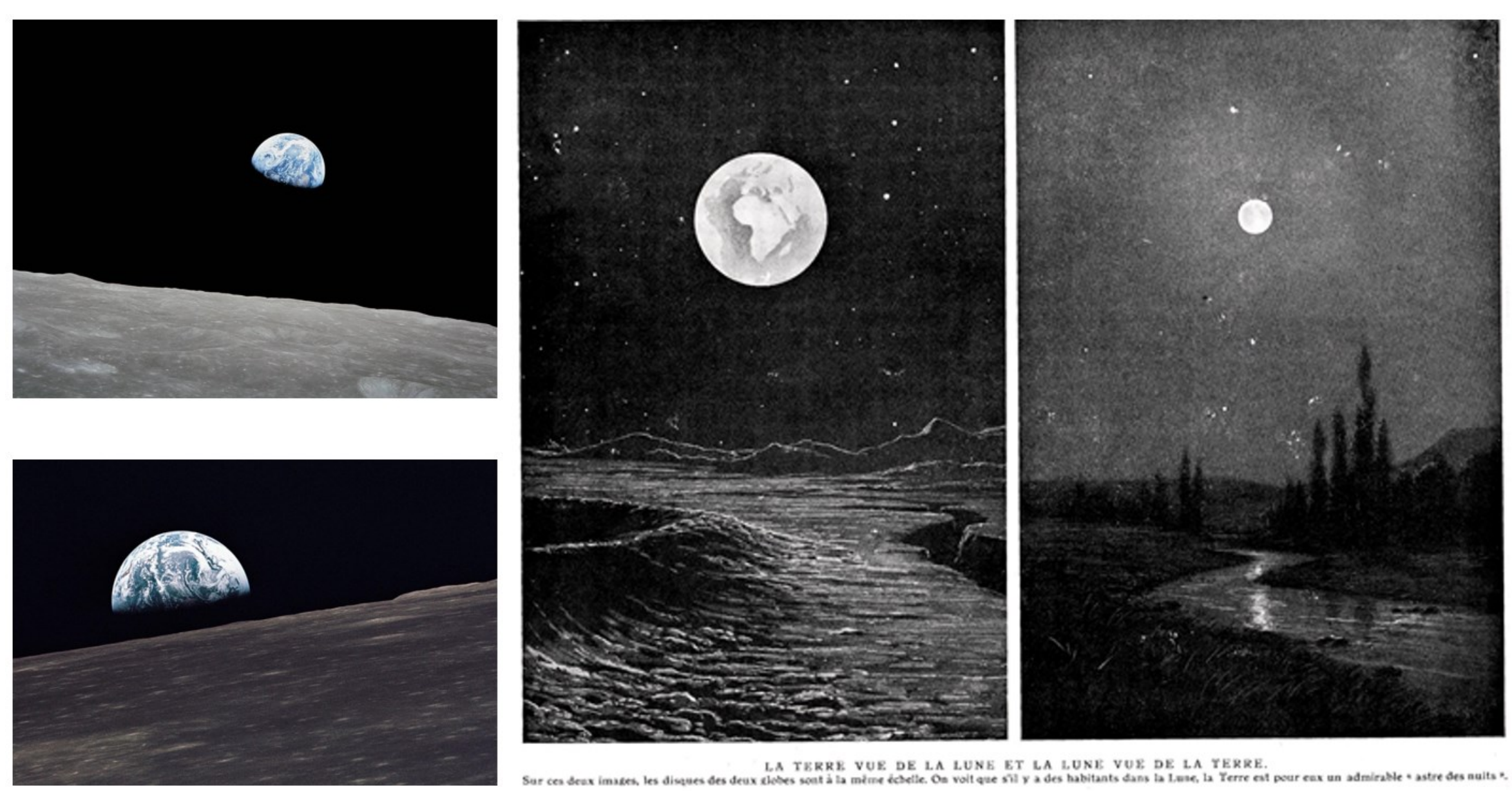

**Figure 5.** Comparative views of Earth and Moon across imagination and observation. Left: Earth above the lunar horizon during the Apollo program (top: Apollo 8 "Earthrise," 1968; bottom: Apollo 17, 1972) [13]. Credit: NASA (public domain). Centre and Right: Earth as seen from the Moon and the Moon as seen from Earth, respectively, from *Le Ciel* (1923) [6], p. 184. (Berget figures: public domain; NASA images: public domain).

## 8. A Modern Reconstruction of the Transit Time reviewed in Le Ciel

The following reconstruction is modern and does not imply that Berget himself used the following Keplerian transfer formalism [11,14].

To reconstruct Berget's second phase, consider a Hohmann-like transfer. A Hohmann transfer is an orbital maneuver used to transfer a spacecraft between two orbits of different altitudes around a central body – in this case the Earth. This maneuver moves the spacecraft from an altitude that is approximately equal to the Earth's radius $r_1 \approx R_\oplus$ to one at the lunar distance, $d$, so that $r_2 \approx d$. The semi-major axis is:

$$a = \frac{r_1 + r_2}{2}. \tag{15}$$

For a minimum-energy Earth–Moon transfer orbit, the time of flight for half the ellipse is:

$$t = \pi \sqrt{\frac{a^3}{\mu_\oplus}}, \tag{16}$$

where $a \approx \frac{6.37\times10^6 + 3.84\times10^8}{2} \approx 1.95 \times 10^8$ m is the semi-major axis of the transfer ellipse and $\mu_\oplus = GM_\oplus = 3.986 \times 10^{14}\ \mathrm{m^3\ s^{-2}}$ is Earth's gravitational parameter. Using these accepted values yields Apollo's one-way travel time $t_{\mathrm{Apollo}} \approx 4.3 \times 10^5\ \mathrm{s} \approx 119\ \mathrm{h} \approx 5$ days. This idealized two-body estimate is not the actual Apollo trajectory. Apollo used a faster trajectory with higher injection energy and mission-specific constraints designed to reduce travel time, rather than the slower minimum-energy Hohmann-like transfer considered here, with typical Earth-to-Moon transit times of approximately 72 h (3 days).

The shorter estimate in *Le Ciel* implies a trajectory with higher specific energy, or larger $v_\infty$, which reduces the transit time. If the spacecraft departs faster than required for a minimum-energy transfer, the time of flight decreases. A simple radial approximation captures this point. Applying conservation of specific mechanical energy (energy per unit mass), gives [12,15]:

$$\frac{1}{2}\left(\frac{dr}{dt}\right)^2 = \frac{GM_\oplus}{r} + \varepsilon, \tag{17}$$

where $\varepsilon$ is the specific orbital energy. For escape trajectories, $\varepsilon > 0$. Rearranging and integrating the differential equation given in (17), the travel time from $r_1$ to $r_2$ is given by

$$t = \int_{r_1}^{r_2} \frac{1}{\sqrt{2\left(\frac{GM_\oplus}{r} + \varepsilon\right)}}\, dr. \tag{18}$$

For sufficiently large $\varepsilon$, over much of the trajectory, $\varepsilon \gg \frac{GM_{\oplus}}{r}$, so that:

$$\frac{GM_{\oplus}}{r} + \varepsilon \approx \varepsilon \tag{19}$$

and then the integral (18) simplifies to:

$$t \approx \int_{r_1}^{r_2} \frac{1}{\sqrt{2\varepsilon}}\, dr = \frac{r_2 - r_1}{\sqrt{2\varepsilon}}. \tag{20}$$

Defining:

$$v_{\infty} = \sqrt{2\varepsilon}, \tag{21}$$

we obtain

$$t \approx \frac{r_2 - r_1}{v_{\infty}}, \tag{22}$$

This is motion at approximately constant speed. Assuming a value of $v_{\infty} \sim 2$–3 km/s, one obtains $t \sim (3.8\times10^{8}\ \mathrm{m}/2.5\times10^{3}\ \mathrm{m.s^{-1}}) \approx 1.5\times10^{5}\ \mathrm{s} \approx 42\ \mathrm{h}$ which is close to Berget's estimate once allowance is made for gravitational deceleration and acceleration along the path. Thus, Berget's value of ≈ 48.5 hours [6] for the coasting phase is consistent with a high-energy, non-minimum transfer trajectory. Berget's value is a physically plausible value, and it is acceptable that it does not match the Apollo value exactly - because real missions include parking orbits, injection geometry, midcourse corrections, lunar targeting, and mission-specific safety constraints.

Qualitatively, what happens is that if the spacecraft leaves Earth with only slightly positive energy, gravity continues to strongly modify the velocity during transit. The spacecraft slows substantially while climbing out of Earth's potential well. Motion is therefore highly non-uniform. In contrast, if the spacecraft has large excess energy, then the gravitational potential term $GM_{\oplus}/r$ quickly becomes small compared with the excess specific energy $\varepsilon$. Gravity then acts only as a perturbation. The speed varies little during the trip, so the motion becomes approximately inertial or "quasi-uniform."

For any of the Apollo lunar missions, the two-body approximation of spacecraft in the field of the Earth (Eqs. (15)–(17)) works remarkably well because its trajectory meant that the motion was relatively fast and direct. Most of the motion can therefore be treated as motion in Earth's gravitational field slightly perturbed by the Moon. The Artemis program, in contrast, operates within the full restricted[2] three-body problem [14,16,17]. However, the transit time to lunar distance remains comparable to Apollo values (about five days), confirming that Berget's estimate pertains to the fundamental Earth–Moon transfer timescale rather than mission-specific architecture. Specifically, Artemis used a low-energy trajectory and not a direct Hohmann-like transfer unlike the fast trajectory/high energy Apollo voyage.

[2] The term "restricted" in the *restricted three-body problem* refers to a simplification of the *full three-body problem* whereby one of the three bodies is assumed to have a negligible mass compared to the two others.

In the Artemis case, the gravitational attraction of both Earth and Moon must be treated simultaneously throughout much of the trajectory

$$\frac{d^2\mathbf{r}}{dt^2} = -\frac{GM_{\oplus}}{r^3}\mathbf{r} - \frac{GM_{\text{Moon}}}{|\,\mathbf{r}-\mathbf{r}_{\text{M}}\,|^3}(\mathbf{r}-\mathbf{r}_M). \tag{23}$$

where $\mathbf{r}_{\text{M}}$ is the position vector of the Moon and $|\mathbf{r}-\mathbf{r}_{\text{M}}|$ is the spacecraft distance from the Moon. The first term represents the gravitational attraction of the Earth while the second represents the attraction of the Moon. The spacecraft therefore moves in a gravitational field that continuously changes in both magnitude and direction. This introduces new richer dynamical structures such as Lagrange points, halo orbits and distant retrograde orbits which are non-existent in the simple two-body problem.

*Lagrange points* are equilibrium points for small mass objects moving under the influence of two larger masses [14]. For example, where the combined gravitational effects of the Earth and Moon balance the orbital motion of a spacecraft, allowing it to remain at a fixed position with respect to the rotating Earth-Moon system. There are five such equilibrium points, three are unstable and two are stable.

*Halo orbits* are three-dimensional looping trajectories around the three unstable Lagrange points which are not simple ellipses. These complex orbits arise from the nonlinear coupling of the Earth and Moon gravitational fields.

The Artemis program also employs *distant retrograde orbits (DROs)*. In such trajectories, the spacecraft follows a large orbit around the Moon whose sense is retrograde relative to the Moon's orbital motion around Earth. These orbits are dynamically unusual because they derive their stability not from the Moon alone, but from the combined Earth–Moon gravitational system. These trajectories are clearly considerably more complex than the classical Keplerian ellipses of Apollo. For their calculation, they require numerical integration, nonlinear dynamical analysis, stability theory, and three-body orbital mechanics [14,16,17]. Yet despite this increased complexity, the fundamental physical sequence identified qualitatively by Berget remains unchanged. Any Earth-to-Moon trajectory must still proceed through the same broad regimes:

1. Escape from Earth's strong gravitational field,
2. Transit through interplanetary space under weakening Earth attraction,
3. Entry into a region where Earth and Moon compete gravitationally,
4. Final capture or controlled motion near the Moon.

Apollo missions realized this sequence through relatively direct Keplerian transfers. The Artemis program realizes it through trajectories arising from a more sophisticated analysis

Therefore, the body with negligible mass does not affect the motion of the two massive bodies, although it remains fully subject to their gravitational field.

of the three-body problem. The mathematical description changes, but the underlying physical structure remains the same.

Berget did not possess the mathematical machinery of modern celestial mechanics. He knew nothing of Lagrange points, halo orbits, or distant retrograde orbits. Nevertheless, he identified, in qualitative pedagogical form, several physically meaningful stages imposed by the Earth–Moon gravitational system itself. Modern astrodynamics places this qualitative picture within a far more rigorous framework, adding orbital geometry, propulsion constraints, trajectory optimization, and multi-body dynamics absent from Berget's account.

## 9. Interplanetary Travel Beyond the Moon

The same chapter of *Le Ciel*, extends space travel to Mars and other planets, thereby projecting Newtonian mechanics far beyond any technological capability of his era. Using scaling arguments based on planetary distances and sustained motion through interplanetary space, Berget concludes that voyages to Mars would require durations measured in several (three) months while that to Venus would take around 47 days, implicitly recognizing the immense scale of the Solar System and the impossibility of rapid transit under classical propulsion. This estimate is remarkably consistent with modern Mars missions. Contemporary spacecraft typically require approximately 6–9 months to reach Mars using Hohmann-type transfer trajectories, depending on launch geometry and mission architecture [11,14,15,18].

## 10. Conclusion: Physics Before Engineering

The forward thinking of early pioneers reviewed in the writing of Berget lies in recognizing that the journey to the Moon is not a single event, but a succession of dynamical regimes governed respectively by Earth's gravity, inertial motion, competing Earth–Moon gravitational fields, and finally lunar gravity. What these pioneers lacked were modern tools of calculation such as angular momentum formalism, propulsion theory, numerical integration, digital computation, and contemporary astrodynamics. However, their insight by taking space-travel seriously enough to estimate the necessary time and phases of the voyage was nevertheless remarkable. The comparison with Jules Verne highlights the transition from literary imagination to physically grounded reasoning.

The interest of the triphasic decomposition summarized in *Le Ciel* lies not in intellectual priority or in quantitative precision, but in its pedagogical clarity and beauty of its exposition. It shows how early twentieth century writers could use elementary Newtonian reasoning to anticipate realistic lunar travel phased into acceleration, coast, transition, and braking stages. Modern mission design is, clearly, *vastly* more complex, requiring orbital mechanics, propulsion theory, guidance, numerical integration, and three-body dynamics. The comparison with Apollo and Artemis is therefore not a claim of direct anticipation, but an instructive historical exercise showing how a simple physical model can foreshadow some qualitative features of later spaceflight.

Berget's discussion is also remarkable for extending beyond celestial mechanics to the human realities of prolonged space travel. In the same chapter, he considers food supply, confinement, duration, and the physiological consequences of long voyages. Berget also recognized that interplanetary space constituted a physically hostile environment and discussed the effects of radiation, heat, and exposure on travelers, although within the framework of early twentieth-century physics rather than the modern understanding of cosmic rays and high-energy particle radiation. Overall, the author has anticipated the general areas of the problems that would later become central to astronautics and human spaceflight. In 1923, when aviation itself was still in an early stage, Berget was already contemplating issues that now dominate planning for lunar and Martian missions, including life-support systems, consumable resources, and the physical and psychological endurance required for interplanetary travel.

Long before the Artemis [18,19] or even Apollo, lunar missions or even the first V2 rockets of World War II, early twentieth century scientists and engineers had already captured a number of essential physical aspects of space travel. They provided reasonable estimates of spaceflight duration, considered extended travel to Mars and other planets, and anticipated the human challenges of prolonged voyages through space. *Le Ciel* stands as a beautiful historical and pedagogical document, reflecting what was known at the time of its publication, and also a historically noteworthy vision of humanity's future beyond Earth. It is fitting to close this article with the somewhat poetic finale of Alphone Berget's *Le Ciel*:

> "Oui, tournons-nous vers le Ciel : il symbolise l'éternelle Vérité; sa contemplation nous élèvera au-dessus des petitesses de la Terre; son étude nous consolera des turpitudes humaines. Et nous en sortirons meilleurs, parce que nous y aurons appris que l'homme, grain de sable dans le double infini de l'Espace et du Temps, ne doit pas s'enorgueillir des découvertes acquises, mais s'efforcer toujours à pénétrer plus avant dans le domaine mystérieux de la Science, dans ce domaine toujours plus vaste à mesure qu'on y progresse davantage, dans ce domaine où l'esprit trouve les seules satisfactions vraiment complètes qu'il nous soit donné de goûter sur cette Terre."

> "*Yes, let us turn toward the Sky: it symbolizes eternal Truth; its contemplation will raise us above the pettiness of the Earth; its study will console us for human turpitudes. And we shall emerge from it improved, because we shall have learned there that man, a grain of sand within the double infinity of Space and Time, must not take pride in discoveries already attained, but must always strive to penetrate further into the mysterious domain of Science–a domain that grows ever larger as one advances within it, a domain in which the mind finds the only truly complete satisfactions granted to us on this Earth.*"

## Acknowledgements

C.F.M. is grateful to Alphonse Berget for, *Le Ciel*, and for the enormous influence this beautiful legacy had, and still has, on him. The financial support of the Natural Sciences and Engineering Council of Canada (NSERC), the Canadian Foundation for Innovation (CFI), the Digital Research Alliance of Canada, and Mount Saint Vincent University is gratefully acknowledged.